\documentclass{article}

\PassOptionsToPackage{numbers, compress}{natbib}

    \usepackage[preprint]{neurips_2023}

\usepackage[utf8]{inputenc} 
\usepackage[T1]{fontenc}    
\usepackage{hyperref}       
\usepackage{url}            
\usepackage{booktabs}       
\usepackage{amsfonts}       
\usepackage{nicefrac}       
\usepackage{microtype}      
\usepackage{xcolor}         
\usepackage{autonum}
\usepackage{graphicx}
\usepackage{cleveref}
\usepackage{xcolor}
\usepackage{xspace}
\usepackage{amsmath}
\usepackage{enumitem}
\usepackage{booktabs}
\usepackage{multirow}
\usepackage{textcomp}
\usepackage{arydshln}
\usepackage{wrapfig}

\DeclareMathOperator*{\argmin}{arg\,min}

\hypersetup{hidelinks}

\title{Fast protein backbone generation \\ with SE(3) flow matching}

\author{%
  Jason Yim\thanks{Work done during an internship at Microsoft Research AI4Science.}$^{\ \ }$\thanks{Massachusetts Institute of Technology}$^{\ \ }$\thanks{Corresponding authors}
    \\
  \texttt{jyim@csail.mit.edu} \\
  \And
  Andrew Campbell$^{* \ddagger}$\thanks{University of Oxford}
  \\
  \texttt{campbell@stats.ox.ac.uk} \\
  \And
  Andrew Y.~K.~Foong$^{\ddagger}$\thanks{Microsoft Research AI4Science}
  \\
  \texttt{andrewfoong@microsoft.com} \\
  \And
  Michael Gastegger$^{\P}$
  \\
  \texttt{mgastegger@microsoft.com} \\
  \And
  José Jiménez-Luna$^{\P}$
  \\
  \texttt{jjimenezluna@microsoft.com} \\
  \And
  Sarah Lewis$^{\P}$
  \\
  \texttt{sarahlewis@microsoft.com} \\
  \And
  Victor Garcia Satorras$^{\P}$
  \\
  \texttt{victorgar@microsoft.com} \\
  \And
  Bastiaan S.~Veeling$^{\P}$
  \\
  \texttt{basveeling@microsoft.com} \\
  \And
  Regina Barzilay$^{\dagger}$
  \\
  \texttt{regina@csail.mit.edu} \\
  \And
  Tommi Jaakkola$^{\dagger}$
  \\
  \texttt{tommi@csail.mit.edu} \\
  \And
  Frank Noé$^{\P}$
  \\
  \texttt{franknoe@microsoft.com} \\
}

\begin{document}

\newcommand{\appendixhead}{
    \textbf{\LARGE Appendix}
}

\newcommand{\TODO}[1]{\textcolor{red}{ {\small \bf (!)} #1}}
\newcommand{\jason}[1]{\textcolor{blue}{ {\small \bf (JY)} #1}}
\newcommand{\edit}[1]{\textcolor{red}{#1}}
\newcommand{\tang}[1]{\mathcal{T}}

\newcommand{\flow}{\phi}
\newcommand{\E}{\mathbb{E}}
\newcommand{\norm}[1]{\left\| #1 \right\|^2 }
\newcommand{\normNoSquare}[1]{\left\| #1 \right\| }

\maketitle

\begin{abstract}
We present \emph{FrameFlow}, a method for fast protein backbone generation using $\textrm{SE}(3)$ flow matching.
Specifically, we adapt FrameDiff, a state-of-the-art diffusion model, to the flow-matching generative modeling paradigm.
We show how flow matching can be applied on $\textrm{SE}(3)$ and propose modifications during training to effectively learn the vector field.
Compared to FrameDiff, FrameFlow requires five times fewer sampling timesteps while achieving two fold better designability. 
The ability to generate high quality protein samples at a fraction of the cost of previous methods paves the way towards more efficient generative models in \textit{de novo} protein design.  
\end{abstract}

\section{Introduction}

Generative models have demonstrated the potential to design novel protein structures for bespoke functions.
Much of this success is due to advancements in diffusion models, which have been applied to various protein representations, ranging from carbon-alpha only \citep{trippe2022diffusion}, to torsion angles \citep{wu2022protein} and the SE(3) backbone frame representation \citep{yim2023se}.
Of these, the frame representation has been shown to achieve state-of-the-art results in \textit{de novo} protein design tasks such as RFdiffusion \citep{watson2023novo}.

However, a major drawback of diffusion models is their inference speed, with $\sim1000$ model forward passes often required to produce high-quality samples.
This can make large-scale inference prohibitively expensive if the score model is large, as in the case of RFdiffusion.
Recently, flow matching methods, which remove stochasticity from the sampling path, have emerged as an alternative to diffusion models, and have been generalised to Riemannian manifolds \citep{lipman2022flow,chen2023riemannian}.
The connection between flow matching and optimal transport is particularly appealing, as the linear interpolating schedule enforces straighter sampling trajectories that can be simulated with fewer integration steps \citep{lipman2022flow}.
These benefits have already been demonstrated in the computer vision domain, where flow matching provides results comparable to diffusion-based models at a fraction of their cost~\citep{pooladian2023multisample}.

Motivated by these results, we develop flow matching in the context of protein backbone generation. We present \emph{FrameFlow}, an adaptation of the FrameDiff \citep{yim2023se} diffusion model to flow matching.
In concurrent work, \citet{bose2023se3stochastic} also develop an $\mathrm{SE}(3)$ flow matching method for protein backbone generation, but don't demonstrate a speed-up during sampling compared to diffusion models.
They focused on using minibatch optimal transport and stochastic differential equations to achieve higher designability.
In contrast, in this work we take advantage of the flow matching framework to focus on improved performance and efficiency.

The paper is organized as follows.
\Cref{sec:method} describes flow matching on the $\mathrm{SE}(3)$ manifold, and introduces FrameFlow for regressing the conditional vector field.
\Cref{sec:experiments} presents our results when training FrameFlow on the SCOPe dataset \citep{chandonia2022scope}.
By using flow matching, we obtain 2 fold better designability, comparable diversity and equal novelty scores compared to FrameDiff, while using five times fewer sampling timesteps.
Compared to GENIE \citep{lin2023generating}, we achieve a 23 times sampling speedup while maintaining a significantly higher designability score.

\section{Method}
\label{sec:method}

\subsection{$\mathrm{SE}(3)$ flow matching}
\label{sec:frame_matching}

\paragraph{Flow matching on Riemannian manifolds.}

Flow matching (FM) \cite{lipman2022flow} is a simulation-free method for learning continuous normalizing flows (CNFs) \cite{chen2018neural}, a class of deep generative models that generates data by integrating an ordinary differential equation (ODE) over a learned vector field. Recently, flow matching has been extended to general Riemannian manifolds \cite{chen2023riemannian}, which we use to model the space of protein backbones. We first give a general introduction to flow matching on manifolds, before specializing to our application.

On a manifold $\mathcal{M}$, the CNF $\flow_t(\cdot): \mathcal{M} \to \mathcal{M}$ is defined by integrating along a time-dependent vector field $v_t(x) \in \mathcal{T}_x \mathcal{M}$ where $\mathcal{T}_x \mathcal{M}$ is the tangent space of the manifold at $x \in \mathcal{M}$:
\begin{equation}
    \frac{d}{dt}\flow_t(x) = v_t(\flow_t(x)), \quad \flow_0(x) = x.
    \label{1}
\end{equation}
Time is parameterized by $t \in [0, 1]$. The flow is used to transform a simple prior density $p_0$ towards the data distribution $p_1$ using the push-forward equation $p_t = [\phi_t]_{*}p_0$, where the density of $p_t$ is
\begin{equation}
    p_t(x) = [\flow_t]_{*} p_0(x) = p_0(\flow_t^{-1}(x)) e^{- \int_{0}^t \text{div}(v_t)(x_s) \, \mathrm{d}s }.
\end{equation}
We refer to the sequence of probability distributions $\{p_t : t \in [0,1] \}$ as the \emph{probability path}. The vector field $v_t$ that generates a given $p_t$ is intractable in general but can be learned efficiently by decomposing the target probability path $p_t$ as a mixture of tractable \emph{conditional} probability paths, $p_t(x | x_1)$. Each conditional path satisfies $p_0(x | x_1) = p_0(x)$, and $p_1(x | x_1) \approx \delta(x - x_1)$. The desired unconditional probability path $p_t$ can then be written as an average of the conditional probability paths with respect to the data distribution: $p_t(x) = \int p_t(x | x_1) p_1(x_1) \, \mathrm{d}x_1$.

Let  $u_t(x | x_1) \in \mathcal{T}_x\mathcal{M}$ be the \emph{conditional vector field} that generates the conditional probability path $p_t(x | x_1)$. 
The key insight of FM is that the unconditional vector field $v_t$ can be learned using an objective which targets the conditional vector field $u_t(x | x_1)$:
\begin{equation}
    \mathcal{L}_{\text{CFM}} := \E_{t, p_1(x_1), p_t(x | x_1)} \left[ \norm{v_t(x) - u_t(x | x_1)}_g \right],
\end{equation}
where $t \sim \mathcal{U}([0, 1])$, $x_1 \sim p_1(x_1)$, $x \sim p_t(x | x_1)$ and $\norm{\cdot}_g$ is the norm induced by the Riemannian metric $g$.
This loss can be reparameterized by defining the conditional flow, $x_t = \psi_t(x_0 | x_1)$, where $\psi_t$ is the solution to $\frac{d}{dt} \psi_t(x) = u_t\left(\psi_t\left(x_0 | x_1\right) | x_1\right)$ with initial condition $\psi_0(x_0 | x_1) = x_0$. The conditional flow matching loss can then be written as:
\begin{align}
    \mathcal{L}_{\text{CFM}} = \E_{t, p_1(x_1), p_0(x_0)} \left[ \norm{v_t(x_t) - \dot{x}_t}_g \right]
\label{9}.
\end{align}
Once trained, samples can be generated by simulating \cref{1} using the learned vector field $v_t$.

\paragraph{Flow matching on $\mathrm{SE}(3)$.}
We now describe the application of FM to protein backbone generation.
The backbone atom positions of each residue in a protein backbone are parameterized by a rigid transformation $T \in \mathrm{SE(3)}$ (see \citet{jumper2021highly,yim2023se}).
Each frame $T = (r, x)$ consists of a rotation $r \in \mathrm{SO}(3)$ and a translation vector $x \in \mathbb{R}^3$.
The protein backbone is made of $N$ residues meaning it can be parameterized by $\mathbf{T} = [T^{(1)}, \dots, T^{(N)}]$ with $\mathbf{T} \in \mathrm{SE}(3)^N$.
Our development focuses on a single frame, but extends to all frames in a backbone since $\mathrm{SE(3)}^N$ is a product space and we choose an additive metric over the frames.
For notational simplicity, we use superscripts to refer to residue indices while subscripts refer to time.

Following~\citet{yim2023se}, we define a metric on $\mathrm{SE}(3)$ by choosing $\langle (a, y), (a', y') \rangle_{\mathrm{SE}(3)} = \langle a, a' \rangle_{\mathrm{SO}(3)} + \langle y, y' \rangle_{\mathbb{R}^3}$ where $\langle a, a' \rangle_{\mathrm{SO}(3)} = \text{Tr}(a a'^{\mathsf{T}})/2$ and $\langle y, y' \rangle_{\mathbb{R}^3} = \sum_{i=1}^3 y_i y'_i$ are the canonical metrics on $\mathrm{SO}(3)$ and $\mathbb{R}^3$ for tangent vectors $a \in \mathfrak{s} \mathfrak{o}(3)$ and $y \in \mathbb{R}^3$, respectively.
This metric enables us to consider $\mathrm{SO}(3)$ and $\mathbb{R}^3$ independently when training and sampling.

Our priors are chosen as the uniform distribution on $\mathrm{SO}(3)$ and the unit Gaussian on $\mathbb{R}^{3}$, $p_0(T_0) = \mathcal{U}(SO(3)) \otimes \mathcal{N}(0, I_3)$.
Following \citet{chen2023riemannian}, the conditional flow $T_t = \psi_t(T_0 | T_1)$ is defined to be along the geodesic path connecting $T_0$ and $T_1$:
\begin{equation}
    T_t = \mathrm{exp}_{T_0} \left(t \mathrm{log}_{T_0}(T_1) \right),
    \label{2}
\end{equation}
where $\mathrm{exp}_{T}$ is the exponential map and $\mathrm{log}_T$ is the logarithmic map at point $T$.
Notably, distance along the geodesic varies linearly with time.
With our choice of metric, \cref{2} simplifies to the following:
\begin{align}
    \text{Translations } (\mathbb{R}^3):& \ x_t = (1-t)x_0 + t x_1 \\
    \text{Rotations } (\mathrm{SO}(3)):& \ r_t = \mathrm{exp}_{r_0} \left(t \mathrm{log}_{r_0}(r_1) \right). \label{6}
\end{align}
Both $\mathbb{R}^3$ and $\mathrm{SO}(3)$ are \emph{simple manifolds} where closed form geodesics can be derived. 
Specifically, $\mathrm{exp}_{r_0}$ can be computed using Rodrigues' formula and $\mathrm{log}_{r_0}$ is similarly easy to compute \citep{yim2023se}.
With these considerations in mind, our overall objective can be written as:
\begin{equation}
    \mathcal{L}_{\mathrm{SE}(3)} = \E_{t, p_1(\mathbf{T}_1), p_0(\mathbf{T}_0)} \left[ \sum_{n=1}^N \left\{ \norm{v_x^{(n)}(\mathbf{T}_t, t) - \dot{x}^{(n)}_t}_{\mathbb{R}^3} + \norm{v_r^{(n)}(\mathbf{T}_t, t) - \dot{r}_t^{(n)}}_{\mathrm{SO}(3)} \right\} \right],
\end{equation}
where $(n)$ refers to the $n$th residue, $t \sim \mathcal{U}([0, 1-\epsilon])$ for $\epsilon=10^{-3}$. The vectors $\{v_x^{(n)}, v_r^{(n)}\}_{n=1}^N$ approximate the vector field as in \cref{9}, and are modeled with an $\mathrm{SE}(3)$-equivariant neural network (\Cref{sec:frameflow}).
Following our definitions of $x_t^{(n)}$ and $r_t^{(n)}$ we compute their time derivatives and approximate them as:
\begin{equation}
    \dot{x}_t^{(n)} =\frac{x_1^{(n)} - x_t^{(n)}}{1-t}, \quad \dot{r}_t^{(n)} = \frac{\mathrm{log}_{r_t^{(n)}}(r_1^{(n)})}{1-t}, \quad v_x^{(n)} :=\frac{\hat{x}_1^{(n)} - x_t^{(n)}}{1-t}, \quad v_r^{(n)} := \frac{\mathrm{log}_{r_t^{(n)}}(\hat{r}_1^{(n)} )}{1-t},
    \label{4}
\end{equation}
where $\{(\hat{x}_1^{(n)}, \hat{r}_1^{(n)})\}_{n=1}^N$ are predictions of the clean frames given the corrupted frames $\mathbf{T}_t$ at time $t$.
Following \citet{yim2023se}, we reparameterize the objective as predicting the clean data:
\begin{align}
    \mathcal{L}_{\mathrm{SE}(3)} = \E_{t, p_1(\mathbf{T}_1), p_0(\mathbf{T}_0)} \Bigg[ \frac{1}{(1-t)^2}\sum_{n=1}^N \Big\{& \norm{\hat{x}_1^{(n)}(\mathbf{T}_t, t) - x_1^{(n)}}_{\mathbb{R}^3} + \\
    &\norm{ \mathrm{log}_{r_t^{(n)}} \left( \hat{r}_1^{(n)} (\mathbf{T}_t, t) \right) - \mathrm{log}_{r_t^{(n)} }\left(r_1^{(n)}\right) }_{\text{SO}(3)}  \Big\} \Bigg].\label{3}
\end{align}

\paragraph{Symmetries.} We perform all modelling within the zero center of mass (CoM) subspace of $\mathbb{R}^{N \times 3}$ as in \cite{yim2023se}. This entails simply subtracting the CoM from the prior sample and all datapoints. As $x_t$ is a linear interpolation between the noise sample and data, $x_t$ will have $0$ CoM also.
This guarantees that the distribution of sampled frames that the model generates is $\mathrm{SE}(3)$-invariant. 
To see this, note that the prior distribution is $\mathrm{SE}(3)$-invariant and the vector field $\{v_x^{(n)}, v_r^{(n)}\}_{n=1}^N$ is equivariant because we use an $\mathrm{SE}(3)$-equivariant architecture to predict $\{\hat{x}_1^{(n)}, \hat{r}_1^{(n)}\}_{n=1}^N$.
Hence by \citet{kohler2020equivariant}, the push-forward of the prior under the flow is invariant.

\subsection{FrameFlow}
\label{sec:frameflow}

\Cref{sec:frame_matching} relies on learning an equivariant vector field using an equivariant neural network. In this section, we discuss the choice of network architecture and additional modifications to improve performance.

\paragraph{Network Architecture.}
To learn $\hat{T}_1^{(n)} = (\hat{r}_1^{(n)}, \hat{x}_1^{(n)})$ for every residue $n$, we utilize the FramePred architecture introduced in FrameDiff \citep{yim2023se} which incorporates Invariant Point Attention (IPA) updates introduced in \citet{jumper2021highly} to encode spatial features and ensure its outputs are equivariant with respect to the input.
Between IPA layers are transformer layers \citep{vaswani2017attention} used to encode sequence-level features.
Unlike FrameDiff, we do not predict the psi angle to recover the oxygen atom but use the planar geometry of the backbone to impute the oxygen atoms, as done in RFdiffusion.
All other hyperparameters, \textit{e.g.} hidden dimensions, and the use of self-conditioning \citep{chen2022analog}, follow FrameDiff.

\paragraph{Loss modifications.} We weight the rotation loss terms in \cref{3} by 0.5 to be on a similar scale as the translation loss.
We notice the loss explodes for $t \approx 1$ due to the $1/(1-t)^2$ term; we found it beneficial to clip this scaling to $1/(1-\text{min}\{t, 0.9\})^2$.

\paragraph{SO(3) inference scheduler.} Our development of $\mathrm{SO}(3)$ FM (\Cref{sec:frame_matching}) follows \citet{chen2023riemannian} in using a linear scheduler $\kappa(t) = 1-t$.
However, we found this schedule to perform poorly for $\mathrm{SO}(3)$ in the context of $\mathrm{SE}(3)$ FM.
Instead, we utilize a exponential scheduler $\kappa(t) = e^{-ct}$ for some constant $c$.
For high $c$, the rotations accelerates towards the data faster than the translations which still follow the linear schedule.
The $\mathrm{SO}(3)$ flow in \cref{6} and vector field in \cref{4} become the following when re-derived,
\begin{align}
    r_t &= \mathrm{exp}_{r_0} \left(\left(1 - e^{-ct} \right) \mathrm{log}_{r_0}(r_1) \right) \\
    v_r^{(n)} &= c\log_{r_t^{(n)}}\left(\hat{r}_1^{(n)}\right).
    \label{7}
\end{align}
We find $c=10$ or $5$ to work well and use $c=10$ in our experiments.
Interestingly, we found the best performance when $\kappa(t) = 1-t$ was used for $\mathrm{SO}(3)$ during training while $\kappa(t) = e^{-ct}$ is used during inference.
We found using $\kappa(t) = e^{-ct}$ during training made training too easy with little learning happening.
The vector field in \cref{7} matches the vector field in FoldFlow when inference annealing is performed.
However, their choice of scaling was attributed to normalizing the predicted vector field rather than the schedule. 

\paragraph{Alternative SO(3) prior.}
Rather than using the $\mathcal{U}(\mathrm{SO}(3))$ prior during training, we find using the $\mathrm{IGSO3}(\sigma=1.5)$ prior \citep{nikolayev1970normal} used in FrameDiff to result in improved performance.
The choice of $\sigma=1.5$ will shift the $r_0$ samples away from $\pi$ where near degenerate solutions can arise in the geodesic.
During sampling, we still use the $\mathcal{U}(\mathrm{SO}(3))$ prior.



\paragraph{Pre-alignment.} Following \citet{klein2023equivariant} and \citet{shaul2023kinetic}, we pre-align samples from the prior and the data by using the Kabsch algorithm to align the noise with the data to remove any global rotation that results in a increased kinetic energy of the ODE.
Specifically, for translation noise $X_0 \sim \mathcal{N}(0, I_3)^N$ 
and data $X_1 \sim p_1$ where $X_0, X_1 \in \mathbb{R}^{N\times 3}$ we solve
$r^* = \argmin_{r \in \mathrm{SO}(3)}\|rX_0 - X_1\|_2^2$ and use the \emph{aligned} noise $r^*X_0$ during training.
We found pre-aligment to aid in training efficiency.

\section{Experiments}
\label{sec:experiments}

\paragraph{Training.} Following GENIE \citep{lin2023generating}, we evaluate FrameFlow by training it on SCOPe with proteins below length 128 for a total of 3938 examples and evaluating on the protein monomer generation task.
Our model is trained for 1 day using two NVIDIA A100-48GB GPUs using the batching strategy from FrameDiff of combining proteins with the same length into the same batch to remove extraneous padding.
We use the Adam \citep{kingma2014adam} optimizer with learning rate 0.0001, $\beta_1 = 0.9, \beta_2 = 0.999$.

\paragraph{Metrics.} To evaluate the model, we sample 10 backbones for every length between 60 and 128 then use ProteinMPNN \citep{dauparas2022protmpnn} to design 8 sequences for each backbone.
We then compute three metrics used in GENIE and FrameDiff: designability, diversity, and novelty.
\textit{Designability} is the main metric where the structure of each of the 8 sequences are predicted using ESMFold \citep{lin2022evolutionary}. Then we compute the minimum RMSD, referred to as \texttt{scRMSD}, between all the ESMFold predictions and the sampled backbone. A sample is deemed designable if \texttt{scRMSD} < 2.0 \AA. Designability is reported as the fraction of designable samples.
\textit{Diversity} is computed by computing the number of structural clusters using MaxCluster \citep{herbertmaxcluster} over all samples with then dividing by the total number of designable samples.
We also report the total number of clusters.
\textit{Novelty} is performed by considering designable samples and using FoldSeek \citep{van2022foldseek} to search for similar structures and computing the highest average TM-score \citep{zhang2005tm} of samples to any chain in PDB, referred to as \texttt{pdbTM}. We report novelty as the average \texttt{pdbTM} across all samples.

\begin{figure*}[!t]
\includegraphics[width=\textwidth]{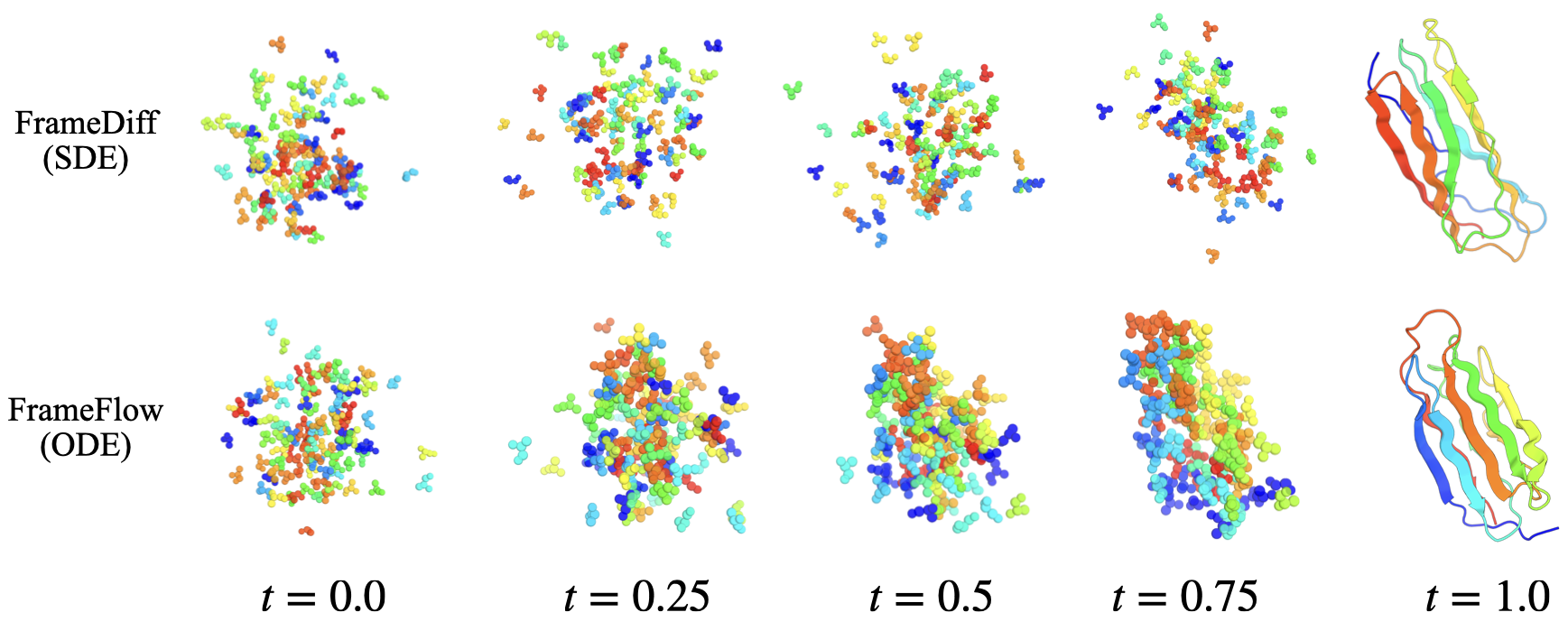}
\centering
\caption{
    Sampling trajectories for FrameFlow (ODE) and FrameDiff (SDE). FrameFlow leads to much straighter integration paths, which leads to structure appearing sooner in the sampling process and allows for fewer timesteps to be used during sampling.
}
\label{fig:traj}
\end{figure*}

\paragraph{Baselines.}
We compare our results to GENIE and FrameDiff, two diffusion models for protein backbones that do not rely on using a pre-trained folding network (unlike RFdiffusion).
We use the GENIE GitHub weights trained on the same training set\footnote{\url{https://github.com/aqlaboratory/genie/tree/main/weights/scope_l_128}} while FrameDiff is re-trained using its default recommended settings.
We expect FrameFlow to underperform RFdiffusion which we were unable to re-train on the smaller dataset.
Our baselines are intended to demonstrate tradeoffs in speed and performance.

\subsection{Results}
\label{sec:results}

We use the Euler-Maruyama integrator for SDE sampling and the Euler integrator for ODE sampling.
We demonstrate the effect of different numbers of integration timesteps for all methods.
Our results are shown in \Cref{5}.
\begin{table}[!ht]
\centering
\caption{Protein backbone generation results.}
\label{5}
\begin{tabular}{llc|ccc}
\toprule
Model & & Timesteps & Designability ($\uparrow$) & Diversity (clusters) ($\uparrow$) & Novelty ($\downarrow$) \\
\midrule
\multirow{3}{*}{GENIE} & \multirow{3}{*}{SDE} & 1000 & 0.22 & 0.76 (131) & 0.54 \\
& & 750 & 0.11 & 0.79 (60) & 0.51 \\
& & 500 & 0.0 & - & - \\
\midrule
\multirow{2}{*}{FrameDiff} & \multirow{2}{*}{SDE} & 500 & 0.42 & 0.36 (104) & 0.66 \\
& & 100 & 0.39 & 0.32 (86) & 0.66 \\
\hdashline
\multirow{2}{*}{FrameDiff} & \multirow{2}{*}{ODE} & 100 & 0.26 & 0.43 (76) & 0.65 \\
& & 10 & 0.16 & 0.53 (59) & 0.66 \\
\midrule
\multirow{3}{*}{\textbf{FrameFlow}} & \multirow{3}{*}{ODE} & 500  & 0.81 & 0.22 (123) & 0.69 \\
& & 100 & 0.77 & 0.28 (147) & 0.67 \\
& & 10 & 0.33 & 0.54 (124) & 0.63 \\
\bottomrule
\end{tabular}
\end{table}
We use SDE sampling for GENIE and FrameDiff since these were the methods used in their respective papers.
In GENIE, we find designability is low while diversity and novelty are favorable compared to FrameDiff and FrameFlow when using 1000 timesteps.
The designability of GENIE even at 1000 timesteps\footnote{GENIE reports 0.85 designability using the scTM > 0.5 criterion. We are able to replicate this finding, but designability in terms of scRMSD is significantly lower.} is significantly lower than that of FrameFlow and FrameDiff at 100 timesteps.
We note that low designability can skew the diversity and novelty metrics since they are defined conditioned on samples being designable.
However, performance in GENIE rapidly deteriorates when we reduce the number of timesteps, and is unusable at 500 timesteps, being unable to produce designable samples.

Importantly, FrameFlow when sampled with only 100 timesteps outperforms the performance of FrameDiff on designability. 
Using the probability ODE sampling procedure for FrameDiff also does not result in improved performance.
FrameFlow's performance deterriorates rapidly with 10 timesteps which other ODE integrators could improve upon.
We note that FrameFlow and FrameDiff use exactly the same architecture. This demonstrates the ability of flow matching to significantly reduce inference costs in protein backbone generation.

Diversity appears lower for FrameFlow. However, this is due to diversity \textit{being inversely proportional to the number of designable samples}.
We follow the diversity definition used in prior works, but note this metric can be artificially high by methods with low designability.
\Cref{5} includes in parantheses the number of clusters which demonstrates FrameFlow discovering more modes than FrameDiff in the data distribution.
GENIE has a high number of clusters and the best novelty indicating its high coverage despite low designability.

While GENIE has less parameters (4.1M) than FrameDiff/FrameFlow (17.4M), it uses expensive triangle updates \cite{jumper2021highly} that requires high memory cost and greater compute for each forward call.
Sampling a length 100 protein with 1000 timesteps on an NVIDIA V100 GPU takes GENIE 128 seconds while for FrameDiff/FrameFlow sampling with 100 timesteps takes 5.7 seconds.

Lastly, we visualize the sampling trajectory of both FrameDiff (SDE) and FrameFlow (ODE) for a length 100 protein in  \Cref{fig:traj}.
Our observations mirror the original motivations behind FM for achieving straighter and faster trajectories.

\section{Discussion}

In this work, we presented $\mathrm{SE}(3)$ flow matching using the previously developed theory from \citet{chen2023riemannian} and \citet{yim2023se}.
We adapted FrameDiff, an $\mathrm{SE}(3)$ diffusion model, into a flow matching model called FrameFlow and demonstrated FrameFlow's superior performance over FrameDiff and GENIE.
Our experiments are preliminary demonstrations of the potential of flow matching to aid in scaling generative models for protein design as neural networks increase in size and complexity.
Concurrent work, FoldFlow, did not exploit the improved speed and efficiency of flow matching, but instead utilized minibatch optimal transport \citep{tong2023improving,pooladian2023multisample} for improved designability.
We believe there is much to explore in the space of flow matching techniques to improve performance in real-world applications with protein design.

\section*{Author contributions}
JY, AYKF, and FN conceived the study.
JY designed and implemented FrameFlow.
JY, AYKF, and AC ran experiments.
JY, AYKF, and AC wrote the manuscript.
MG, JJL, SL, VGS, and BSV contributed to the codebase used for experimentation and are ordered alphabetically.
RB, TJ, and FN offered supervision.
FN advised and oversaw the study.




\bibliography{references}

\end{document}